\documentclass[3p,times]{elsarticle}

\usepackage{ecrc}

\usepackage{epstopdf}

\volume{00}

\firstpage{1}

\journalname{Nuclear Physics A}

\runauth{Author1 et al.}

\jid{nupha}

\jnltitlelogo{Nuclear Physics A}

\usepackage{graphicx}
\usepackage{amsmath,amssymb}

\begin{document}

\begin{frontmatter}



\title{Theory of Jet Quenching in Ultra-Relativistic Nuclear Collisions}

\author{Guang-You Qin}
\address{Institute of Particle Physics and Key Laboratory of Quark and Lepton Physics (MOE), Central China Normal University, Wuhan, 430079, China}




\begin{abstract}

We present a short overview of recent progress in the theory of jet quenching in ultra-relativistic nuclear collisions,
including phenomenological studies of jet quenching at RHIC and the LHC, development in NLO perburbative QCD calculation of jet broadening and energy loss, full jet evolution and modification, medium response to jet transport, and lattice QCD and AdS/CFT studies of jet quenching.

\end{abstract}


\end{frontmatter}



\section{Introduction}
\label{intro}

One of main goals of ultra-relativistic nuclear collisions, such as those performed at the Relativistic Heavy-Ion Collider (RHIC) and the Large Hadron Collider (LHC), is to create a novel state of matter, called quark-gluon plasma (QGP), and study its various properties.
Large transverse momentum quarks and gluons, produced from early stage hard scatterings, have been regarded as very useful probes of such highly excited nuclear matter.
These hard partonic jets interact with medium constituents during their propagation through the QGP medium before fragmenting into hadrons.
The interaction with medium usually causes partons to lose energy, therefore observables associated with jets are modified as compared to the vacuum jets such as the case in elementary nucleon-nucleon collisions.

The basic framework for studying jet production in ultra-relativistic nuclear collisions is perburbative QCD factorization paradigm, i.e., processes involving large transverse momentum ($p_T$) transfer may be factorized into long-distance and short-distance pieces.
For example, the cross section of single inclusive high $p_T$ hadron production in elementary nucleon-nucleon collisions may be obtained as follows:
\begin{eqnarray}
d\sigma_h \approx \sum_{abjd} f(x_a) \otimes f(x_b) \otimes d\sigma_{ab\to jd} \otimes D_{j\to h}(z_j).
\end{eqnarray}
In the above factorized formula, $f(x_a)$, $f(x_b)$ are parton distributions functions (PDF) and $D(z_j)$ is fragmentation functions (FF).
These non-perturbative long-distance quantities are universal and usually obtained from global fitting to various experimental measurements, such as $e^+ e^-$ and deep-inelastic scattering (DIS) experiments, etc.
The partonic scattering cross sections $d\sigma_{ab\to jd}$ are short-distance quantities involving large $p_T$ transfer, and thus may be calculated using perturbative QCD techniques.

In phenomenological studies of jet energy loss and jet quenching ultra-relativistic heavy-ion collisions, the above formula needs some modification due to the presence of hot and dense QGP:
\begin{eqnarray}
d\tilde{\sigma}_h \approx \sum_{abjj'd} f(x_a) \otimes f(x_b) \otimes d\sigma_{ab\to jd} \otimes P_{j\to j'} \otimes D_{j' \to h}(z_j').
\end{eqnarray}
The additional piece $P_{j \to j'}$ is to take into account the interaction between the hard partons $j$ and the QGP medium before fragmenting into high $p_T$ hadrons.
Often one combines parton-medium interaction function $P_{j \to j'}$ and vacuum fragmentation function to define so-called medium-modified fragmentation function $\tilde{D}_{j \to h} \approx \sum_{j'} P_{j\to j'} \otimes D_{j' \to h}$; one may also define medium-modified parton cross section by combining parton-medium interaction with vacuum cross section.
Although the above factorized formula have been widely used in phenomenological studies of jet modification and energy loss, no formal proof of factorization is available yet for ultra-relativistic heavy-ion collisions.

\section{Radiative and Collisional Jet Energy Loss}

In the study of jet quenching in ultra-relativistic heavy-ion collisions, two different medium-induced mechanisms are usually considered for jet energy loss: elastic collisions with medium constituents and induced bremsstrahlung processes.
The energy loss occurred in binary elastic collisions is usually referred to as collisional jet energy loss.
The scatterings between hard partonic jets with medium constituents usually induce additional radiation which carry away part of the parent parton's energy; such mechanism is called radiative energy loss.

During last decades, much effort has been focused on studying radiative jet energy loss. A number of approaches have been developed in the literature.
Based on the assumptions made in different formalisms when calculating single gluon emission spectrum, they may be cast into two categories: multiple soft scatterings, such as Baier-Dokshitzer-Mueller-Peigne-Schiff-Zakharov (BDMPS-Z), \cite{Baier:1996kr, Zakharov:1996fv}, Amesto-Salgado-Wiedemann (ASW) \cite{Wiedemann:2000za} and Arnold-Moore-Yaffe (AMY) \cite{Arnold:2001ba} formalisms, versus few hard scatterings, such as Gyulassy-Levai-Vitev (GLV) \cite{Gyulassy:1999zd} and higher twist (HT) \cite{Wang:2001ifa} formalisms.
Recently, various improvements have been done to the original jet energy loss formalisms.
For example, AMY formalism has been developed to include finite medium length effect \cite{CaronHuot:2010bp}.
GLV formalism has been extended to finite dynamical medium, and recently the effect of non-zero magnetic mass was also included \cite{Djordjevic:2008iz, Djordjevic:2011dd}.
HT formalism has been extended to incorporate multiple scatterings \cite{Majumder:2009ge}.

In addition to the difference in single gluon emission spectrum, different evolution schemes have been used when calculating multiple gluon emissions.
One popular method is Poisson convolution which assumes that multiple gluon emissions are independent.
This method has been widely used in BDMPS, ASW and GLV formalisms, to get the probability distribution $P(\Delta E)$ of parton energy loss \cite{Salgado:2003gb, Renk:2006sx, Wicks:2005gt}:
\begin{eqnarray}
P(\Delta E) = \sum_{n=0}^{\infty} \frac{e^{-\langle N_g \rangle}}{n!} \left[ \prod_{i=1}^{n} \int d\omega \frac{dI(\omega)}{d\omega} \right] \delta\left( \Delta E - \sum_{i=1}^{n} \omega_i \right),
\end{eqnarray}
where $dI/d\omega$ is the spectrum for single gluon emission.
In AMY formalism, the coupled rate equations have been used to obtain the time evolution of quark and gluon jet momentum distribution $f(p)=dN(p)/dp$ \cite{Jeon:2003gi, Qin:2007rn}. The rate equations can be schematically written as:
\begin{eqnarray}
\frac{df(p,t)}{dt} = \int dk \left[ f(p+k,t) \frac{d\Gamma(p+k,k,t)}{dkdt} - f(p,t) \frac{d\Gamma(p,k,t)}{dkdt} \right],
\end{eqnarray}
where $d\Gamma(p,k,t)/dkdt$ is the rate for a parton with momentum $p$ to lose momentum $k$.
HT formalism is based on perturbative QCD power expansion; it solves DGLAP-like evolution equations to obtain the medium-modified fragmentation function $\tilde{D}(z,Q^2)$ \cite{Majumder:2011uk, Qin:2009gw}:
\begin{eqnarray}
\frac{\partial \tilde{D}(z,Q^2)}{\partial\ln Q^2} = \frac{\alpha_s}{2\pi} \int \frac{dy}{y} P(y) \int d\zeta^- K(\zeta^-, Q^2) \tilde{D}(z/y, Q^2),
\end{eqnarray}
where $K(\zeta^-, Q^2)$ is the kernel for parton-medium scattering which induces additional radiation in the medium.
A systematic comparison of different jet energy loss models has been performed in the framework of a ``brick" of QGP in Ref. \cite{Armesto:2011ht}.

Collisional jet eneergy loss was first studied by Bjorken in 1982 \cite{Bjorken:1982tu}.
Compared to radiative component of jet energy loss, collisions energy loss is usually considered to be small for light flavor (leading) partons, especially when jet energy is sufficiently large.
But in realistic calculation of nuclear suppression factor $R_{AA}$ at RHIC and the LHC kinematics, collisional energy loss may give sizable contribution \cite{Wicks:2005gt,Qin:2007rn, Schenke:2009ik}.
The contribution of collisional energy loss is more prominent for heavy flavor partons.
Taking bottom quarks as a example, it has been shown in Ref. \cite{Cao:2013ita} that below around $15$ GeV collisional energy loss dominates while  at high energy radiative component dominates.
Collisional energy loss is also a very important ingredient when studying full jet evolution and energy loss \cite{Qin:2010mn}, and the response of medium to jet transport \cite{Qin:2009uh,Neufeld:2009ep} as will be discussed Sec. (\ref{full_jet},\ref{medium_response}).

\section{Phenomenological Studies at RHIC and the LHC}

The main purpose of jet quenching study is to figure out the detailed interaction mechanisms between jets and the QGP medium, and to improve our knowledge of hot and dense nuclear matter.
One important way to achieve this is to perform systematic studies of jet quenching observables and compare to a wealth of available experimental measurements.
Various phenomenological studies have been performed in the literature for a wealth of jet quenching observables, such as the suppression of single inclusive high $p_T$ hadron production, dihadron suppression and photon-hadron correlations in relativistic nucleus-nucleus collisions \cite{Bass:2008rv,Armesto:2009zi,Chen:2010te,Zhang:2007ja,Renk:2008xq,Qin:2009bk}.
Recently, much interest has been shifted to the quantitative extraction of various jet transport coefficients.
These transport coefficients can generally be written as the correlations of gluon fields, thus may provide much insight into the internal structure and the properties of QGP matter that hard jets traverse.
One of the most important jet quenching parameters is $\hat{q}$ \cite{Baier:1996kr},
\begin{eqnarray}
\hat{q} = \frac{1}{L} \int \frac{d^2k}{(2\pi)^2} k_\perp^2 P(\vec{k}_\perp,L)
\approx \frac{4\pi\alpha_s C_s}{N_c^2 -1}\int dy^-  \langle F^{\mu +}(0) F_{\mu}^+(y^-) \rangle,
\end{eqnarray}
where $P(\vec{k}_\perp)$ is the probability distribution of transverse momentum transfer between the propagating partons and medium.
This parameter not only quantifies the transverse momentum broadening of the jet \cite{Majumder:2007hx,Qin:2012fua}, but also controls the size of medium-induced radiative energy loss.
As has been pointed out in Ref. \cite{Majumder:2007zh}, the precise determination of $\hat{q}/T^3$ combined with the knowledge of shear viscosity over entropy ratio $\eta/s$ can provide important clues to our understanding of QGP properties, e.g., when and at what scale a weakly-coupled quark gluon system at sufficiently high temperatures changes to a strongly-coupled fluid at RHIC and the LHC energies.

\begin{figure}
\begin{center}
\includegraphics*[width=9.cm]{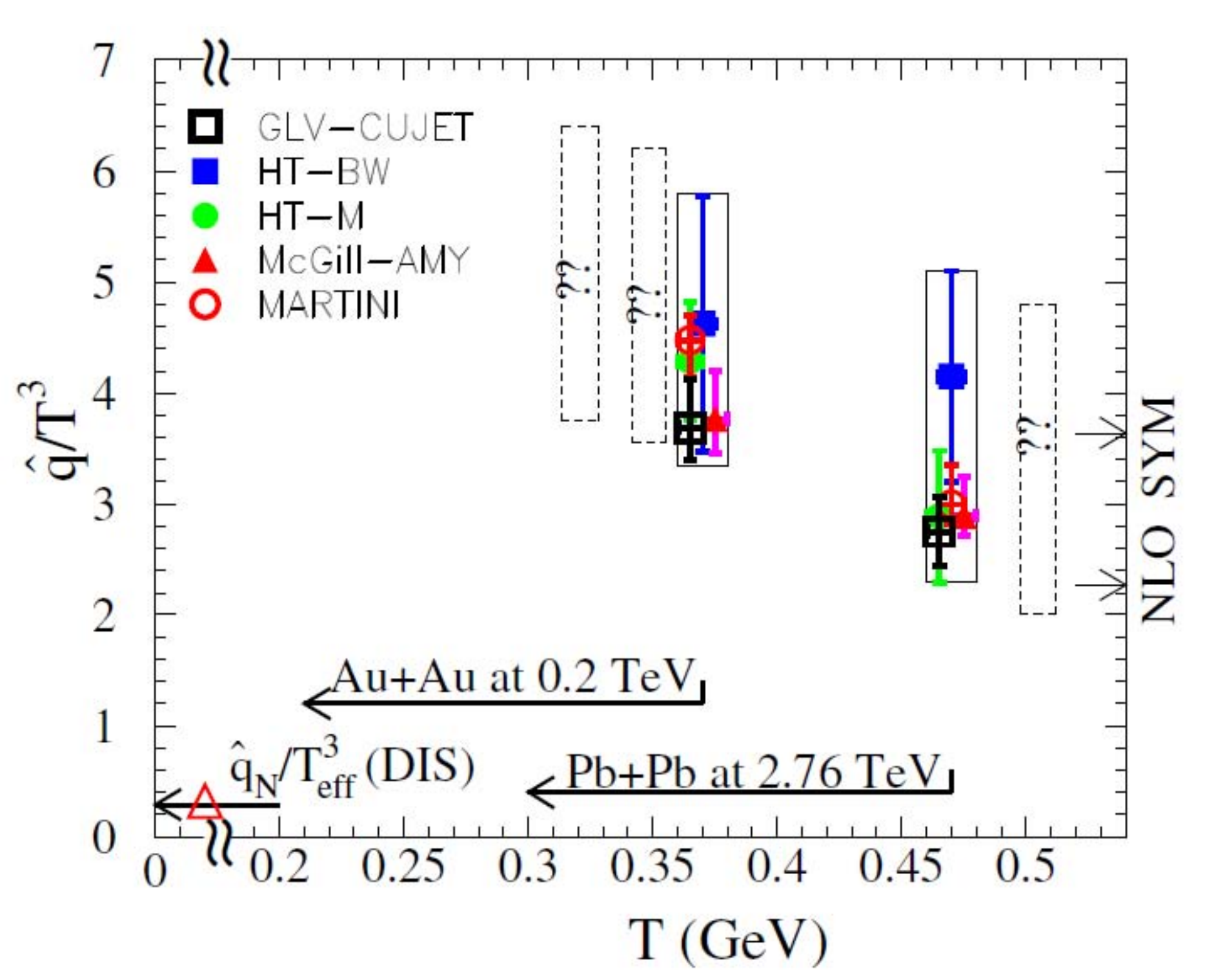}
\caption{(Color online) The extracted values of scaled jet transport parameter $\hat{q}/T^3$ by using single inclusive hadron suppression factor $R_{AA}$ at both RHIC and LHC.
The shown values are for a quark jet with initial energy of $10$~GeV at the center of the most central A-A collisions and at an initial proper time $\tau_0 = 0.6$~fm/c.
The figure is taken from Ref. \cite{Burke:2013yra} done by JET Collaboration.
}
\label{JET_Collaboration}
\end{center}
\end{figure}

Very recently a collaborative effort have been performed to quantitatively extract jet quenching parameter $\hat{q}$ within the framework of JET Collaboration \cite{Burke:2013yra}
by a few jet quenching groups: McGill-AMY \cite{Qin:2007rn}, Martini-AMY \cite{Young:2011ug}, HT-M \cite{Majumder:2011uk}, HT-BW \cite{Chen:2011vt}, DGLV-CUJET \cite{Xu:2014ica}.
In this work, the space-time evolutions of the bulk QGP matter produced in heavy-ion collisions at RHIC and the LHC are described utilizing realistic viscous hydrodynamics simulations.
By comparing the calculations from five different energy loss models to experimental measurements, the values of jet quenching parameter $\hat{q}$ have been  extracted.

One important result from such collaborative effort is shown in Fig. \ref{JET_Collaboration}.
The range of $\hat{q}$ values as constrained by the measured single hadron nuclear modification factors $R_{AA}$ at both RHIC and LHC are obtained as:
\begin{eqnarray}
\frac{\hat{q}}{C_s T^3} = \left\{
                                \begin{array}{ll}
                                  3.5 \pm 0.9, & \hbox{$T \approx 370$ {\rm MeV (at RHIC)},} \\
                                  2.8 \pm 1.1, & \hbox{$T \approx 470$ {\rm MeV (at the LHC)}.}
                                \end{array}
                              \right.
\end{eqnarray}
This translates into $\hat{q} \approx$ 1.2-1.9~GeV$^2$~fm for a quark jet at the highest temperatures reached in the most central Au+Au collisions at RHIC and Pb+Pb collisions at LHC.

One may clearly see that the scaled dimensionless quantity $\hat{q}/T^3$ is temperature dependent.
In the future, it is of great importance to map out the temperature dependence of jet quenching parameter by extending the current study to the future higher energy Pb-Pb collisions at the LHC and lower energy collisions at RHIC.
The expected values of $\hat{q}$ in A-A collisions at other collision energies (0.063~ATeV, 0.130~ATeV and 5.5~ATeV) are shown by dashed boxes.
We can see that the values of $\hat{q}$ in hot QGP matter are much higher than those of cold nuclei (the value of $\hat{q}_N/T_{\rm eff}^3$ in cold nuclei constrained from DIS experiments is indicated by the triangle in the figure).
The result from next-to-leading order (NLO) Super-Yang-Mills (SYM) calculation \cite{Zhang:2012jd} is also shown by two arrows on the right axis of the figure.

We also note some other recent phenomenological studies of jet quenching observables at RHIC and the LHC energies, such as those based on DGLV formalism \cite{Djordjevic:2011dd}, $dE/dx$  model \cite{Betz:2014cza}, and soft-collinear effective theory (SCET) \cite{Kang:2014xsa}, etc.
In the future, we may include more studies in the JET collaboration framework to fully take into account the systematic uncertainties.
Also more experimental observables should be utilized to get tighter constraints on jet energy loss models and jet quenching parameters.

\section{Developments in Next-to-Leading Order Calculations}

In phenomenological studies of jet quenching in ultra-relativistic heavy-ion collisions, leading-order formalisms of jet energy loss have been applied so far.
In Ref. \cite{Armesto:2011ht}, it is found that much of model difference can be attributed to specific approximations made in the model calculations, (such as eikonal, soft and collinear approximations.
To better constrain the models, it is important to investigate the effects of different approximations or to relax these approximations when building up jet quenching formalisms.
One such effort was performed in Ref. \cite{Abir:2013ph}, where the effect of non-eikonal large angle scatterings is studied and it is found to be small for partons with energies larger than 10-15~GeV.
In Ref. \cite{Apolinario:2014xla}, the medium-induced radiation is re-studied by going beyond eikonal approximation, and the finite energy effect has been taken into account.

To achieve a systematic estimation of the errors made in leading-order calculation, we need build a fully next-to-leading order (NLO) framework for calculating jet quenching and jet energy loss.
The first effort along this direction was performed in Ref. \cite{Wu:2011kc, Liou:2013qya}, where NLO correction to parton transverse momentum broadening was studied in the framework of BDMPS formalism.
The correction due to medium-induced radiation is found to be sizable; in particular parton transverse momentum broadening receives a double logarithmic contribution $\ln^2(L/l_0)$, with $L$ the length of the medium and $l_0$ the scale associated with medium constituents.
Recently, it was argued in Ref. \cite{Iancu:2014kga, Blaizot:2014bha} that such double logarithmic correction may be absorbed by a redefinition/renormalization of jet quenching parameter $\hat{q}$.
In Ref. \cite{Iancu:2014kga, Blaizot:2014bha}, the scale evolution of $\hat{q}$ was further studied.
It is found that when the medium size $L$ is large, parton transverse momentum broadening as well as parton energy loss receive additional medium-length dependence, $\Delta E \propto \hat{q}_0 L^{2+\gamma}$ with $\gamma = 2\sqrt{\alpha_s N_c/\pi}$, as compared to traditional BDMPS jet energy loss formalism.

The renormalization of jet quenching parameter $\hat{q}$ has also been studied in Ref. \cite{Kang:2013raa}, where NLO QCD corrections to the transverse momentum broadening are computed in the framework of hadron production semi-inclusive deep-inelastic scattering and lepton pair production in p-A collisions.
In this work, a factorization formula is derived for $k_\perp^2$-weighted cross section; the collinear divergence is absorbed into the redefinitioon of nonpertubative PDF and twist-4 quark-gluon correlation functions.
The factorization leads to a DGLAP evolution equation for parton distribution function, as well as a new QCD evolution equation for twist-4 quark-gluon correlation functions.
From this new QCD evolution equation, the scale $\mu$ dependence of jet quenching parameter $\hat{q}(\mu^2)$ may be obtained if the momentum and spatial correlations of two nucleons inside the nucleus are neglected.

\section{Full Jet Evolution and Modification}
\label{full_jet}

In recent years, full jets have been extensively studied in ultra-relativistic heavy-ion collisions.
With the large kinematics available at the LHC, we now have the opportunity to investigate medium effects on jets with transverse energies over a hundred GeV.
The basic idea of reconstructing full jets is to recombine final state hadron fragments and to infer the information about the original hard partons and the medium effect on them.
Since both leading and sub-leading fragments are included, full jets are expected to provide more discriminative power than leading hadron observables \cite{Vitev:2009rd}.
One big challenge in studying full jets in relativistic heavy-ion collision as compared to elementary collisions is the contamination from the fluctuating hydrodynamic background.
Sophisticated experimental techniques have been developed and utilized to disentangle jets from fluctuating background.

The first jet measurement from the LHC heavy-ion program was the momentum imbalance/asymmmetry of correlated back-to-back jet pairs \cite{Aad:2010bu,Chatrchyan:2011sx}.
We observe a strong modification of dijet momentum imbalance distribution in Pb-Pb collisions as compared to p-p collisions at the LHC, while the the angular distribution is largely unchanged.
Similar observations have been obtained for full jets correlated with high $p_T$ direct photons in Pb-Pb collisions at the LHC \cite{Chatrchyan:2012gt}.
These results indicate that the away-side subleading jets experience significant amount of energy loss when propagating through the produced QGP.
Various model calculations based on jet energy loss have been performed to explain the observed dijet and $\gamma$-jet momentum imbalance \cite{Qin:2010mn,CasalderreySolana:2010eh,Lokhtin:2011qq,Young:2011qx,He:2011pd,Renk:2012cx,Zapp:2012ak,Blaizot:2013vha,Dai:2012am,Qin:2012gp,Wang:2013cia}.

\begin{figure}
\begin{center}
\includegraphics*[width=12.cm,height=4.cm]{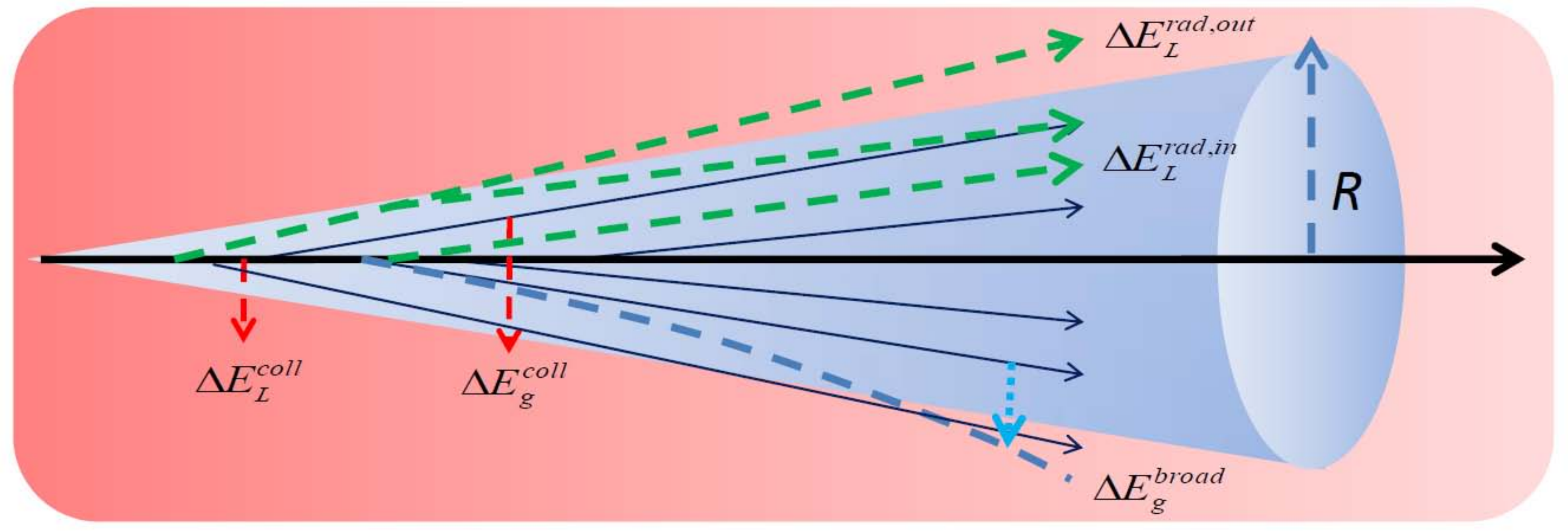}
\caption{(Color online) A schematic illustration of the evolution of full jet, and different medium-induced processes that contribute to full jet energy loss and modification in a quark-gluon plasma.
}
\label{full_jet_evolution}
\end{center}
\end{figure}

The evolution of full jets in QGP medium is schematically illustrated in Fig. \ref{full_jet_evolution}, where the thick solid arrow through the center of jet cone represents the leading parton of the jet, and other lines represent the accompanying gluons.
Compared to leading hadron observables, a few additional ingredients need be taken into account when studying full jets in QGP medium.
The radiated gluons may lose energy or get deflected by interacting with the medium constituents; some of the gluons may be kicked out of the jet cone.
These ingredients can be encoded by solving the following Boltzmann transport equation \cite{Qin:2010mn}:
\begin{eqnarray}
\frac{df_g(\omega, k_\perp, t)}{dt} = -\hat{e}\frac{\partial f_g}{\partial \omega} + \frac{1}{4} \hat{q} \nabla_{k_\perp}^2 f_g + \frac{dN_g^{\rm rad}}{d\omega dk_\perp^2 dt},
\end{eqnarray}
where $f_g(\omega, k_\perp, t)$ is the momentum distribution of the accompanying gluons of the full jets.
In the equation, the first and second terms represent the collisional energy loss and transverse momentum broadening due to interacting with medium constituents, and the last term denotes the gluon radiation induced by the medium.

By solving the above evolution equation, we may obtain the information of full jets after propagating through the medium.
In particular, the energy of the original full jet may be decomposed as:
\begin{eqnarray}
E_{\rm jet} = E_{\rm in} + E_{\rm lost} = E_{\rm in, rad} + E_{\rm out, rad} + E_{\rm out, brd} + E_{\rm th, coll}.
\end{eqnarray}
Interestingly, the last term, i.e., the medium absorption or the thermalization of the soft radiations, is found to give the largest contribution to full jet energy loss.
This is not a surprising result since the medium modification of soft components of the jet or accompanying gluons at large angles are expected to easier than the inner hard core of the jet.
The picture is often referred to as jet collimation: soft partons of the jet are stripped off when they propagate through the hot and dense nuclear medium \cite{CasalderreySolana:2010eh}.

Similar results have also been obtained in Ref. \cite{Blaizot:2013vha, Armesto:2011ir, CasalderreySolana:2012ef, Mehtar-Tani:2014yea} where it is argued that if medium color field varies over jet transverse size, the shower partons of the full jets will lose color coherence due to their interaction with the surrounding medium.
Such color decoherence effect will open up the phase space for soft and large angle gluon radiation, as compared to traditional BDMPS formalism.
Utilizing rate equation to study multiple gluons emissions, it is found that the energy of jets can be rapidly degraded into many soft gluons which carry energies of medium temperature.
In terms of radiated gluons, there are three different phase spaces separated by two scales, $x_0$ and $x_{\rm th}$.
The total energy of the original jet may be decomposed as:
\begin{eqnarray}
E_{\rm jet} = E_{\rm in} + E_{\rm lost} = E_{\rm in}(x>x_0) + E_{\rm out}(x_{\rm th}<x<x_0) + E_{\rm flow}(x<x_0).
\end{eqnarray}
The radiated gluons with $x>x_0$ are inside the jet cone and thus part of final reconstructed jet.
For $x < x_0$, the radiation is outside jet cone. The radiation with $x<x_{\rm th}$ is soft and thus quickly thermalized and flows into the medium.

The substructures or fragmentation profiles are also of great interest in full jet study as they may provide additional information about jet-medium interaction \cite{Vitev:2009rd}.
Both longitudinal and transverse jet fragmentation profiles have been measured ATLAS and CMS Collaborations for Pb-Pb collisions at the LHC  \cite{Chatrchyan:2012gw, Chatrchyan:2013kwa, Aad:2014wha}.
For transverse directions, little change is observed at small radius $r$ compared to p-p collisions, while there is a significant excess at large $r$ accompanied by a depletion at intermediate $r$.
For the momentum (fraction) distribution of jet fragments,  there is an excess at both small and large $z (p_T)$, and a depletion in the fragmentation profiles (functions) is observed at intermediate $z (p_T)$.

There are some challenges and complications in the study of full jet observables due to their sensitivity to various details of jet-medium interaction and experimental setups.
For example, different treatments of recoiled partons may lead to different pictures of jet energy loss, and may results in different jet internal structures in the final state \cite{Wang:2013cia}.
In Ref. \cite{Ma:2013gga}, it has been shown that fragmentation and recombination mechanisms may lead to different jet fragmentation profiles.
It is also very important to take into account all experimental jet finding conditions when comparing model calculations of jet substructures with experimental measurements \cite{Ramos:2014mba}.

\section{Medium Response to Jet Transport}
\label{medium_response}

Jets lose energy during their propagating through QGP medium; some of the lost energy is deposited into the medium and make medium excitations.
One way to study the medium response to jet propagation is to solve the following hydrodynamic equation,
\begin{eqnarray}
\partial_{\mu} T^{\mu \nu}(x) = J^{\nu}(x)
\end{eqnarray}
where the source term represents the jet energy and momentum deposition profiles which may be calculated from jet quenching calculations.

The study of medium response and collective excitations is very interesting and important since it may provide a direct probe to the speed of sound $c_s$ of the QGP medium.
Also it is a necessary ingredient of jet-medium interaction and helpful for understanding many observables associated with jets, e.g., where the lost energy manifests in the final state.
In Ref. \cite{Tachibana:2014lja}, the medium response to the lost energy from two back-to-back partons is studied using a (3+1)-dimentional ideal hydrodynamics, and the redistribution of the lost energy after hydrodynamic evolution is also calculated.
The finding from this study is qualitatively consistent with the measurements by CMS Collaboration, i.e., the lost energy from the jet are carried by soft particles at large angles \cite{Chatrchyan:2011sx}.

It should be noted that the energy deposition profiles for full jets are quite different from single partons due to the fact that the radiated partons may serve as additional sources depositing energy and momentum into medium and significantly increase the length dependence of energy deposition rate \cite{Qin:2009uh,Neufeld:2009ep}.
Jet energy deposition profiles are also sensitive to some details in the model calculations, such as the cutoff energy often applied to determine which part of radiation is treated as thermalized and flows to the medium \cite{Qin:2009uh,Neufeld:2009ep,Renk:2013pua,Floerchinger:2014yqa}.
The spatial distribution of the jet energy/momentum deposition profiles is also important for studying the response of medium to jet transport.
For example, it is found in Ref. \cite{Neufeld:2011yh} that collinear radiation may produce very nice cone-like structure of medium response, but a wide spatial distribution of energy deposition profiles expected from large angle radiation may destroy such cone-like structure.

\section{Jet Quenching from Lattice QCD and AdS/CFT}

While perturbative QCD based jet energy models have been very successful in studying jet quenching in relativistic heavy-ion collisions, there exist many model discrepancies due to specific approximations applied in model \cite{Armesto:2011ht, Burke:2013yra}.
Lattice QCD calculation can provide some guideline and constraints on our jet quenching studies, e.g., the values of various jet transport coefficients that are typically obtained from phenomenological calculations.
In Ref. \cite{Majumder:2012sh}, jet quenching parameter $\hat{q}$ was first computed within the framework finite temperature lattice gauge theory.
The calculation was carried out in quenched $SU(2)$, and extrapolated to the case of $SU(3)$ with $2$ flavors of quarks, yielding a value of $\hat{q}$=1.3-3.3~GeV$^2$/fm for a gluon jet at a temperature $T=400$~MeV.
Recently within the framework of a dimensionally reduced effective theory (electrostatic QCD), the contribution from soft QGP modes to jet quenching parameter $\hat{q}$ was evaluated numerically, and a value of $\hat{q}$=6~GeV$^2$/fm is obtained for RHIC energies \cite{Panero:2013pla}.
A first principle calculation of collisional kernel $C(k_\perp)$ was also carried out in Ref. \cite{Laine:2013apa} and the shape of the collision kernel $k_\perp^3 C(k_\perp)$ is found to be consistent with Gaussian at small $k_\perp$.

AdS/CFT correspondence provides another non-perturbative approach and reference for studying jet quenching in a strongly-coupled plasma.
A recent study utilizing more realistic shooting string description of energetic quarks could achieve better agreement with experiment data \cite{Ficnar:2013qxa}.
A NLO calculation which takes into account the corrections from finite t'Hooft coupling produces a value of jet quenching parameter $\hat{q}$ much closer to these extracted from perturbative QCD-based phenomenological studies.
The substructure of full jets was studied in Ref. \cite{Chesler:2014jva}, where it is found that jets emerging from the plasma look almost like vacuum jets, but with reduced energy and larger opening angle.

\section{Summary}

We have provided a short review of the recent development in theoretical and phenomenological studies of jet quenching in ultra-relativistic heavy-ion collisions.
While significant progresses have been achieved in various directions, such as the quantitative extraction of jet quenching parameter, full jet energy loss and substructures, medium response to jet transport, and so on, there are still many open questions and a lot to be improved.
More detailed phenomenological studies are needed in order to map out the temperature dependence of various jet transport coefficients.
A fully next-to-leading order framework for studying jet energy loss and modification is still unavailable.
It is important to build a framework which allows for simultaneous simulation of jet transport and medium response by combining realistic hydrodynamic models and jet energy loss/deposition calculations.
The detailed thermalization mechanisms for jet deposited energy/momentum is not fully understood.
Complications exist in both theoretical and experimental studies of full jets and their detailed substructures.
The effort along these and other directions will further improve our knowledge of jet-medium interaction, and help us to achieve more comprehensive understanding of the novel properties of hot and dense QGP produced in high energy nuclear collisions.

\section{Acknowledgments}
This work was supported in part by the Natural Science Foundation of China under grant no 11375072.











\bibliographystyle{h-physrev5}
\bibliography{GYQ_refs}


\end{document}